\documentclass[10pt,journal,compsoc]{IEEEtran}

\usepackage[numbers]{natbib}
\usepackage{hyperref}
\hypersetup{
  colorlinks = true, 
  citecolor = blue, 
  linkcolor = blue, 
  urlcolor = black
}
\usepackage{graphicx}
\usepackage{amsmath}
\usepackage{amssymb}
\usepackage{amsfonts}
\usepackage{mathtools}
\usepackage{pifont}
\usepackage{xcolor}
\usepackage{float}
\usepackage{url}
\usepackage{tabularx}
\usepackage{balance}
\usepackage{comment}
\usepackage{multirow}
\usepackage{amsthm}
\usepackage{ragged2e}
\usepackage{algorithm}
\usepackage{algorithmicx}
\usepackage{algpseudocode}
\usepackage{setspace}
\usepackage[compact]{titlesec}
\newcommand{\ct}[1]{\textcolor{black}{#1}}

\newcommand{\dt}[1]{\textcolor{black}{#1}}
\newcommand{\rw}[1]{\textcolor{black}{#1}}
\newcommand{\mr}[1]{\textcolor{black}{#1}}
\newcommand{\D}[1]{\begin{normalsize}\begin{tt}#1\end{tt}\end{normalsize}}

\newtheorem{theorem}{Theorem}

\begin{document}

\IEEEoverridecommandlockouts
\IEEEpubid{\begin{tabular}[t]{@{}l@{}}\copyright 2022 IEEE. Personal use of this material is permitted. Permission from IEEE must be obtained for all other uses, in any current or future media, including\\reprinting/republishing this material for advertising or promotional purposes, creating new collective works, for resale or redistribution to servers or lists, or\\reuse of any copyrighted component of this work in other works\end{tabular}}


\title{PackCache: \ct{An Online} Cost-driven Data Caching Algorithm in the Cloud}


\author{Jiashu~Wu,
        Hao~Dai,
        Yang~Wang\thanks{* This work is supported by Key-Area Research and Development Program of Guangdong Province (2021B0101400005), Yang Wang is the corresponding author (yang.wang1@siat.ac.cn).},
		Yong~Zhang,
        Dong~Huang,
        and~Chengzhong~Xu,~\IEEEmembership{Fellow,~IEEE}
\IEEEcompsocitemizethanks{
  \IEEEcompsocthanksitem Jiashu Wu, Hao Dai and Dong Huang are with Shenzhen Institute of Advanced Technology, Chinese Academy of Sciences, Shenzhen, 518055, and the University of Chinese Academy of Sciences, Beijing, 100049. 
  \IEEEcompsocthanksitem Yang Wang and Yong Zhang is with Shenzhen Institute of Advanced Technology, Chinese Academy of Sciences, Shenzhen, 518055. 
  \IEEEcompsocthanksitem Chengzhong Xu is with the University of Macau, Macau, 999078. 
  \IEEEcompsocthanksitem DOI: 10.1109/TC.2022.3191969
  }
\thanks{Manuscript received Jan 00, 2022; revised Jan 00, 2022.}}

\markboth{IEEE Transactions on Computers}%
{Shell \MakeLowercase{\textit{Wu et al.}}: PackCache: An Online Cost-driven Data Caching Algorithm in the Cloud}


\IEEEtitleabstractindextext{%
\begin{abstract}
In this paper, we study a data caching problem in the cloud environment, where multiple frequently co-utilised data items could be packed as a single item being transferred to serve a sequence of data requests dynamically with reduced cost. To this end, we propose an online algorithm with respect to a homogeneous cost model, called \emph{PackCache}, that can leverage the FP-Tree technique to mine those frequently co-utilised data items for packing whereby the incoming requests could be cost-effectively served online by exploiting the concept of anticipatory caching. We show the algorithm is $2/\alpha$ competitive, reaching the lower bound of the competitive ratio for any deterministic online algorithm on the studied caching problem, and also time and space efficient to serve the requests. Finally, we evaluate the performance of the algorithm via experimental studies to show its actual cost-effectiveness and scalability.
\end{abstract}


}

\maketitle
\IEEEpubidadjcol

\IEEEdisplaynontitleabstractindextext

\IEEEpeerreviewmaketitle


\IEEEraisesectionheading{\section{Introduction}\label{sec:introduction}}

\IEEEPARstart{A}{s} the complexity of applications keeps increasing, \mr{various tasks in real life may require more than one data items to collaboratively complete the task. }For instance, during text preprocessing \cite{anandarajan2019text_nlp}, a tokenisation model and a token dictionary may be required together for Chinese sentence analysis. Hence, given that data items are sometimes correlated in accesses, they thus can be packed \ct{to serve the requests via caching at reduced cost}. As a way to reduce communication overhead and minimise response delay for data accesses in the context of cloud computing, data caching is playing an essential role in managing data in such applications. Despite of the co-utilisation pattern of data items, past works seldom transfer data items via caching in a packed manner \cite{wang2017data,wang2013data}, \mr{especially in online scenarios, where data layout and access optimisation is infeasible due to the online nature. Hence, it hinders the cost-efficiency of applications in the cloud environment when serving data requests. }

Motivated by the observation that data-item co-utilisation is frequent in the cloud environment, and the cost benefit \ct{can be} brought by the packing mechanism, an offline caching algorithm, which combines a greedy strategy and a dynamic programming technique to serve requests in a packable manner, was proposed to minimise the service cost \cite{huang2019dp_greedy}. However, the offline scenario is not always realistic. \ct{On the other hand, the Jaccard Similarity-based (JS-based) method used in \cite{huang2019dp_greedy} might} also perform unstably in online settings to mine the frequent co-utilised data items \cite{bury2019similarity_jaccard_sensitive}, \ct{thus crippling the deployments of the algorithm in practice.}

In this paper we address these issues by presenting an online packable data caching algorithm, \ct{called} \emph{PackCache}, which utilises the FP-Tree to discover frequently co-utilised data items as it performs more stable and generates less variations \cite{bury2019similarity_jaccard_sensitive} \ct{than the JS-based method} in online settings. Additionally, the algorithm leverages the concept of anticipatory caching \cite{wang2021} to serve the requests and maintain the data caches in an online fashion. \ct{To our best knowledge, the work in this paper is among the first to tackle the caching problem in an online packable manner. }

The \emph{PackCache} algorithm is designed by following the setting adopted in our previous studies \cite{wang2017data,veeravalli2003network,wang2013data}, where a homogeneous cost model is employed \cite{Mansouri2017} --- for each cache server, the storage cost is fixed and for each pair of servers, the communication costs are also identical. Based on this model, we further show that the \emph{PackCache} algorithm is $2/\alpha$ competitive, where $\alpha$ \mr{is} the discount factor, i.e., the cost saving ratio achieved by the packing mechanism. Moreover, we also prove that $2/\alpha$ is the lower bound of the competitive ratio of the packable caching problem for any deterministic online algorithm, which verifies that the proposed algorithm is tight in terms of competitive performance. Finally, the \emph{PackCache} algorithm is both time and space efficient in serving a request in \dt{$O(1)$ time} and $O(n^2)$ space, where $n$ is the length of the request sequence. 

To evaluate its actual performance in reality, we also implement the \emph{PackCache} algorithm and conduct experiments to show its cost-effectiveness and scalability. 
In summary, we make the following contributions in this paper: 
\vspace{-0.3cm}
\begin{itemize}
	
	\item \ct{We study} a cost-driven data caching problem with packable serving mechanism under the homogeneous cost model and propose a cost-efficient \emph{PackCache} algorithm that utilises the FP-Tree to discover frequently co-utilised data items and adopt the anticipatory caching mechanism to serve requests. 
	
	\item \ct{We show} that the \emph{PackCache} algorithm achieves a competitive ratio of $2/\alpha$, which is also shown to be the lower bound of the online packable caching problem for any deterministic online algorithm. 
	
	\item \ct{We implement} the \emph{PackCache} algorithm and evaluate its performance in practice. The results show that compare with its individually-served counterpart, the \emph{PackCache} algorithm is cost effective and scalable, \rw{i.e., under $\alpha$ value of $0.8$ and $0.6$, a cost reduction of $5\%$ and $10.7\%$ can be achieved. }
	
\end{itemize}
\vspace{-0.3cm}

The rest of the paper is organised as follows: Section \ref{sec:related_work} presents some related works and compares them with ours to demonstrate our research opportunities. After that, we present our model and the detailed problem formulation in Section \ref{sec:problem_formulation}. The \emph{PackCache} algorithm, its competitive ratio analysis and complexity analysis are presented in Section \ref{sec:an_online_2_approximation_algorithm}. Experimental setups and results are illustrated in Section \ref{sec:simulation_studies}. The last section concludes the paper.


\section{Related Work}
\label{sec:related_work}

The caching problems can be divided into online case and offline case distinguished by different ways to receive requests. In the offline case, the full knowledge about requests is available in advance, while the online case knows nothing about the future requests sequence. In the offline setting and a QoS perspective, Zhang et al., \cite{8169053} studied the delay-optimal cooperative caching in the edge environment and proposed a greedy caching placement algorithm with an approximation ratio of $(1 - 1/e)$ in linear time. By following a similar idea, Zhang et al., \cite{8422371} introduced a collaborative hierarchical caching mechanism with an attempt to maximise the overall cache hitting rate. George et al., \cite{7873661} unified both goals by utilising a cooperative caching algorithm with proactive cache updating policy to reduce the delay in online video access and jointly improve the cache hit ratio. 

However, our work differs from previous works in terms of system model and problem goal. In particular, instead of being capacity-oriented and aiming to maximise the cache hit ratio, the goal of our algorithm is to minimise the cost of request serving given adequate resources in the cloud. 

Later, in the offline setting and a cost perspective, Li et al., \cite{LI2021153} established a Markov model and a multiple linear regression model to assist caching usage prediction and guide the caching placement. 
From the theoretical perspective, \rw{Khanafer et al., \cite{6566944} formulated the computation and caching cost trade-off as a constrained ski-rental problem assisted by the first or second moment of the arrival distribution, which outperformed existing approaches in worst-case competitive ratio. }
\rw{Puttaswamy et al, \cite{10.1145/2168836.2168845} and Erradi et al., \cite{ERRADI2020110457} also tackled the storage cost saving by utilising a hybrid adaptive storage solution under multiple storage services, and via a tier-wise object placement algorithm, respectively. Both methods benefited the cost saving during cloud data storage. }
\rw{Wang et al., \cite{wang2021} proposed a 2-competitive online algorithm that applied an individually-served manner to handle the incoming requests and utilised anticipatory caching to maintain the caches of data items under the content delivery edge network. }However, none of the above works attempted the packable mechanism during cloud caching. 

Recently, Huang et al., \cite{huang2019dp_greedy} introduced the DP\_Greedy algorithm that combined an existing dynamic programming (DP)-based algorithm and a greedy strategy to effectively cache data items in the cloud environment with packing being enabled. However, their work was only feasible in offline settings, leaving the online setting untouched. 

All aforementioned methods \ct{are either in absence of packing mechanism when serving data requests at all or only feasible in offline settings that possess complete knowledge about requests, such as} the work in \cite{huang2019dp_greedy}. \ct{To the best of our knowledge}, \ct{there is still in shortage of cost-effective packable data caching algorithms working in online settings}. \ct{Our work instead fills this gap in a realistic and cost-effective way by incorporating the co-relationships between data items into an online algorithm via a packing mechanism to serve data requests.}


\section{Problem Formulation}
\label{sec:problem_formulation}

In this section, we describe the problem formulation of the cost-driven packable data caching problem in details. We first define some useful concepts that will be used in this paper, and then give a standard form of the solution to the problem following the idea proposed in study \cite{wang2017data,huang2019dp_greedy}. 

\subsection{Problem Model}
\label{sec:system_model}

Suppose in a cloud environment, there are $k$ distinct data items with diverse co-utilisation patterns. The set of data items are denoted by $D = \{d_{1}, d_{2}, \cdots, d_{k}\}$, which will be cached in a fully connected network with $m$ cache servers, denoted by $S = \{s_{1},s_{2}, \cdots, s_{m}\}$. A sequence of data requests, $R = \{r_{1}, r_{2}, \cdots, r_{n}\}$, are made to request these data items, where the tuple $r_{i} = <s_{j}, t_{i}, D_{i}>$ represents that request $r_{i}$ is made at server $s_{j}$ ($s_{j} \in S$) at time $t_{i}$ for a data item subset $D_{i}$. For each request $r_{i}$, it can either request a single data item, i.e., $D_{i} = \{d_{i1}\}, d_{i1} \subseteq D$, or can request two data items, i.e., $D_{i} = \{d_{i1}, d_{i2}\}, d_{i1}, d_{i2} \in D$. 

When serving a data request, the shared data items need to be either held locally in the cache of server that receives the data request, or be replicated and transferred from another server to the server with the received request to satisfy the request. After being used, the data will be destroyed at certain time to achieve minimal caching cost so that the requests can be served in a cost-efficient manner. 

Different from previous works, in this paper we consider the online setting, i.e., there is no knowledge of where and when each data request is made and which data item subset is requested. For simplicity, we assume that there exists at most one request per time instance as many other previous studies assumed \cite{wang2017data,veeravalli2003network,wang2013data,huang2019dp_greedy}, so we can use $t_{i}$ to represent the arrival time of the request $r_{i}$. 

As the functionalities of cloud applications become more sophisticated, it is likely that two correlated data items could be frequently utilised together (i.e., co-utilised) by requests. In this circumstance, packing these data items as a package to serve data requests jointly is both convenient and cost-effective. We define a discount factor $\alpha$, $(0 < \alpha \leq 1)$, which measures the ratio between cost of serving a request of two frequent data items in a packed manner, and the cost of the individually-served non-packing case. Note that the packing will not be leveraged during caching for the online case to avoid time-consuming unpacking during single data requests. On the other hand, for infrequently co-utilised data items, the packing mechanism will not be applied. Since if data items are frequently co-utilised, the application that requests these data items usually possesses an optimised way to handle and unpack the package, hence the burden caused by unpacking is negligible. While for infrequent data item pairs, the lack of optimised unpacking mechanism and subsequent single data accesses yield extra cost burdens. Hence, it is not worth packing infrequent data items. In this paper, we only consider packages with two data items, but without loss of generality, the algorithm is convenient to be extended to multiple data item packing cases. 

\begin{figure}[!ht]
\vspace{-7mm}
	\setlength{\abovecaptionskip}{0pt}
	\setlength{\belowcaptionskip}{-15pt}
	\centering
	\includegraphics[scale=0.28]{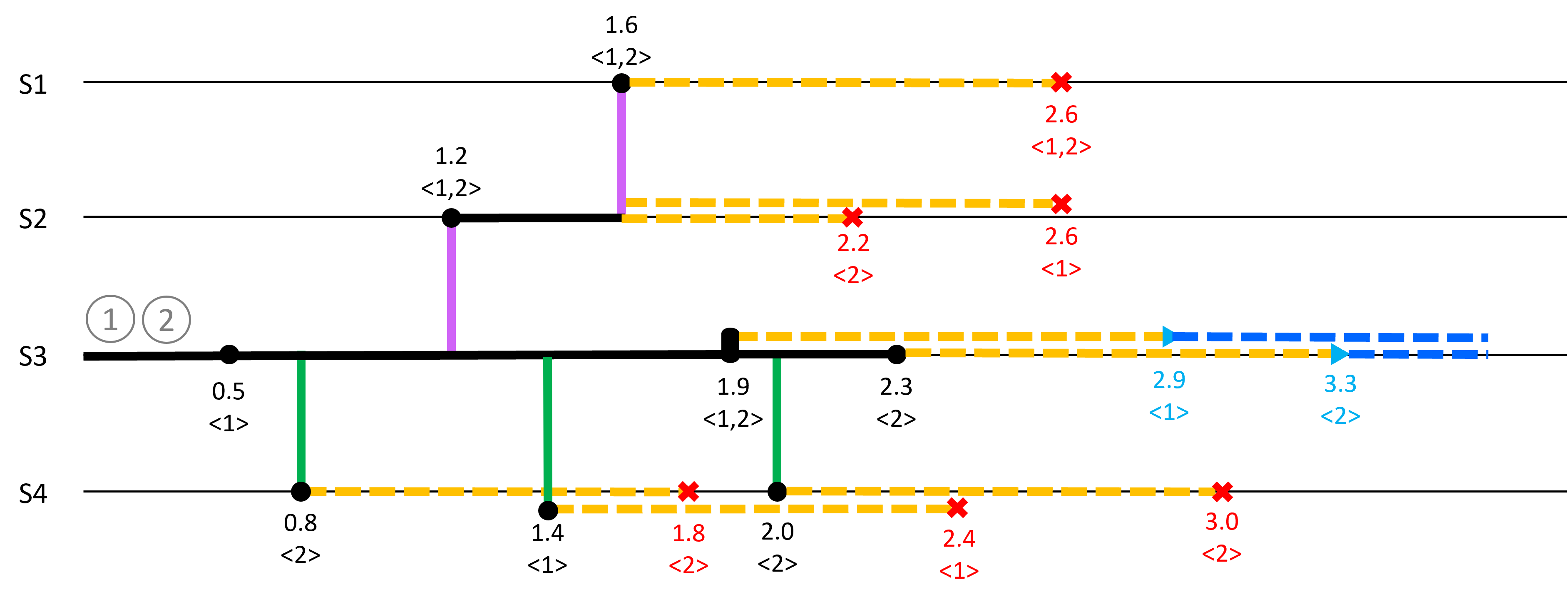}
	\caption{A feasible schedule (bold lines) of a packable caching model ($\mu = \lambda$). The data items are initially on server S$3$. Each vertex and numbers below it represent at time $t$ a request is asking for data item in $<>$. Horizontal lines represent data caching, while green and purple vertical lines indicate individual and packed data item transferring, respectively. The dashed lines indicates extra caching made by the system. The red cross represents cache elimination, while the blue rectangle represents although the expiration is detected, the data is still kept to prevent data loss. The cost is $C = (1 + 2 + 4.7 + 3) \mu + (3 + 2 * 2 * \alpha) \lambda$. }
	\label{Fig2}\vspace{-2mm}
\end{figure}

We also adopt the space-time diagram \cite{huang2019dp_greedy} to clarify the problem. A feasible schedule (shown in Fig.~\ref{Fig2}), is a way to use caching or transferring to get all data requests satisfied along the timeline, and a standard form of a schedule is that all transfers occur at the request time instance. The work \cite{veeravalli2003network} confirmed that there exists at least one optimal schedule which belongs to the standard form. Different from previous research \cite{wang2017data}, we consider the multiple data item caching problem and we take data item's co-utilisation into account and enable the packing mechanism, instead of serving them individually. Besides, our work stands out from \cite{huang2019dp_greedy} as our proposed algorithm works in the online setting, which makes no assumption regarding data request sequence, and provide the corresponding competitive ratio with its lower bound. 

In terms of request satisfaction, since there exists a sequence of data requests, we define the request being satisfied as the data items in that specific request are satisfied. 

\subsection{Cost Model}
\label{sec:cost_model}

\ct{We adopt the same model presented in \cite{huang2019dp_greedy} to conduct our study.} A universal caching costs per time unit for each server is applied, denoted by $\mu$, and the transfer costs between any pair of servers are also identical, denoted by $\lambda$. Suppose $C_{ij}^{p}, (1 \leqslant p \leqslant k)$ represents the cost to serve a data item $d_p$ contained in request $r_{j} = <s_{j}, t_{j}, D_{j}>$, i.e., $d_p \in D_{j}$, and request $r_{i}=<s_{i}, t_{i},D_{i}>$, $d_p \in D_{i}$, is the most recent request for data item $d_p$ before $r_j$.  If $r_j$ is the first request for data $d_p$, $t_i = 0$. We give a formal definition of $C_{ij}^{p}$ as follows:
\begin{equation} 
	C_{ij}^{p}=
	\left\{
	\begin{array}{lr}
	(t_{j}-t_{i})\mu+\epsilon \lambda, & t_{j} > t_{i}  \\
	+\infty,& \mbox{Otherwise} 
	\end{array}
	\right.
\end{equation}
where $\epsilon$ is a variable with value in $\{0, 1\}$ to signify how to treat the transfer cost during the computation of $C_{ij}^{p}$. Specifically, if the request $r_{j}$ is served by a local data cache resulted from $r_{i}$, i.e., $s_{i} = s_{j}$, then $\epsilon=0$. Otherwise, $\epsilon=1$, implying the requested data item $d_p$ is first cached from $t_{i}$ to $t_{j}$ on cache server $s_{i}$, and then transferred from $s_{i}$ to $s_{j}$ to serve request $r_{j}$. If $d_p$ is one of the data item in a request for a frequent data pair and server $s_i$ contains the requested data pair, the corresponding transfer cost $\lambda$ will have the discount factor $\alpha$ being applied, i.e., $\epsilon \alpha \lambda$. \rw{Note that the cost model is generalisable to reflect the overheads faced by the algorithm. Considering that in practice the cost of transferring a unit of data is fixed, then the cost reduction proportionally reflects the reduction in terms of data communication overhead. Similarly, considering that the network condition provided by modern cloud service providers are relatively stable, then the communication overhead reduction also proportionally translates to the decrease of time overhead. }

\subsection{Problem Goal}
\label{sec:problem_goal}

To satisfy a data request, data items may need to be either cached locally to serve the subsequent requests made on this server or transferred from another cache server that has the requested data item so that the request could be satisfied. A transfer operation implies the data item is replicated, then the copy is transferred and cached in the destination server and then destroyed in the future for cost saving. Since the replication cost and deletion cost are always constants, they can be merged into the transfer cost or the caching cost. Without loss of accuracy, we assume these cost are free as \mr{in many previous studies} \cite{wang2017data,veeravalli2003network,wang2013data,huang2019dp_greedy}. 

The problem goal is to find an optimal schedule among many feasible schedules so that the total cost to serve all these data requests is minimised. We denote $\Pi(i)$ as all feasible schedules to satisfy the requests up to $r_{i}$, $\phi(R)$ \ct{as} a feasible schedule and $\phi^{*}(R)$ \ct{as the optimal} for this $n$-length request sequence, i.e., $|R| = n$. 
Each schedule $\phi(R)$ has a cost $cost(\phi(R))$, which is defined as follows: 
\begin{equation}
	cost(\phi(R))=\sum_{r_{i}\in R}cost(r_{i})=\sum_{r_{i}\in R}\sum_{d_{p}\in D_{i}}cost(d_{p})
\end{equation}
whereby we have the same formal definition of the problem goal as in \cite{huang2019dp_greedy}:
\begin{equation}
\phi^{*}(R)=\mathop{\arg\min}_{\phi(R) \in \Pi(R)} \ \ cost(\phi(R))
\end{equation}

In this paper, we study the online form of this problem and give an $2/\alpha$-competitive online algorithm. 


\section{An Online $2/\alpha$-Competitive Algorithm}
\label{sec:an_online_2_approximation_algorithm}

Given that the full information regarding data requests are usually not available in advance in practice, to make the algorithm more generalisable, we propose the \emph{PackCache} algorithm that can work under an online manner. In this section, we first describe the workflow and rationale of the proposed \emph{PackCache} algorithm, which is a $2/\alpha$-competitive algorithm. Then, we prove the competitive ratio and its lower bound. We finally analyse the time and space complexity of the \emph{PackCache} algorithm. 

\makeatletter
\newenvironment{breakablealgorithm}
{
	\begin{center}
		\refstepcounter{algorithm}
		\hrule height.8pt depth0pt \kern2pt
		\renewcommand{\caption}[2][\relax]{
			{\raggedright\textbf{\ALG@name~\thealgorithm} ##2\par}%
			\ifx\relax##1\relax 
			\addcontentsline{loa}{algorithm}{\protect\numberline{\thealgorithm}##2}%
			\else 
			\addcontentsline{loa}{algorithm}{\protect\numberline{\thealgorithm}##1}%
			\fi
			\kern2pt\hrule\kern2pt
		}
	}{
		\kern2pt\hrule\relax
	\end{center}
}
\makeatother
\begin{breakablealgorithm}\label{alg:online_packcache_algorithm}
\caption{The \emph{PackCache} algorithm}
\justifying
	\begin{algorithmic}[1]
	    \small
		\Require online request in the form of $r_{i}=<s_{j}, t_{i}, D_{i}>$
		\Ensure the average cost to serve this request sequence, denoted by ave\_cost
		\State /* all data items are initially located at cache server $s_{1}$ */
		\State \textbf{Initialise}: $c[d_k] \leftarrow 1$; $E[d_k]^j \leftarrow 0$, $1 \le j \le m$
		\State transfer cost $C_T \leftarrow 0$, caching cost $C_C \leftarrow 0$
		\If{(request $r_i$ arrives $s^j$ at time $t_i$)}
			\State \dt{use $frequent\_item\_miner(r_i)$ to mine frequently} co-utilised data items based on minimum support $\gamma$ using request history up to time $t_i$
			\State $C_T \leftarrow C_T + serve\_request(r_i)$
		\EndIf
		\If{(a copy $d_p$ expires on $s_j$ at $t_i$)}
			\State $copy\_expire(d_p, s_j, t_i)$
		\EndIf
		\State \Return average request cost
	\end{algorithmic}
\end{breakablealgorithm}

In the \emph{PackCache} algorithm, we maintain a global counter $c[d_k]$, which keeps track of the number of alive copies of data item $d_k$. Without loss of generality, we assume that all data items are initially cached at server $s_1$ with only one copy, i.e., $c[d_k] = 1$. Besides, we also maintain a local expiration time tracker $E[d_k]^j$ which stands for the expiration time of data item $d_k$ on server $s_j$ and is initialised to be $0$. A previous request recorder $r_{<k>}^j$ is also maintained to record the latest request of $d_{k}$ on server $s_j$. The initialisation process has been shown in \D{line 2 - 3} in Algorithm \ref{alg:online_packcache_algorithm}. 

The \emph{PackCache} algorithm constitutes $3$ components, namely the \textit{FP-Tree-based frequent data itemset miner} as shown in Algorithm \ref{alg:frequent_miner}, the \textit{Request serving component} as shown in Algorithm \ref{alg:serve_request} and the \textit{Data item copy expiration handler} as shown in Algorithm \ref{alg:copy_expire}. 

\begin{breakablealgorithm}\label{alg:frequent_miner}
	\caption{Function $frequent\_item\_miner(r_i)$}
	\justifying
		\begin{algorithmic}[1]
		    \small
			\Require an incoming request in the form $r_{i}=<s_{j}, t_{i}, D_{i}>$
			\Ensure the $FreqI$, i.e., the frequent data itemsets
			\If{(request $r_i$ is a double data item request)}
				\State add request $r_i$ into request history
				\State update FP-Tree
			\EndIf
			\State retrieve all data item pairs that have their support greater than or equal to the minimum support $\gamma$ to form $FreqI$
			\State \Return $FreqI$
		\end{algorithmic}
\end{breakablealgorithm}

As shown in Algorithm \ref{alg:frequent_miner} and \D{line 5} in Algorithm \ref{alg:online_packcache_algorithm}, upon receiving a new data request, the \emph{PackCache} algorithm utilises FP-Tree to discover frequently co-utilised data itemsets with a given minimum support $\gamma$ based on the request history available till the current time $t_i$. As indicated in \D{line 1 - 3} in Algorithm \ref{alg:frequent_miner}, only double data item requests will be added into the request history to guide the mining process. The advantage of utilising the FP-Tree-based frequent itemset miner over the Jaccard Similarity-based one is that the FP-Tree-based frequent itemset miner performs more stable when data requests come in an online manner, and hence it will generate less variations \cite{bury2019similarity_jaccard_sensitive,moulton2018maximally_jaccard_sensitive}. 

\begin{breakablealgorithm}\label{alg:serve_request}
	\caption{Function $serve\_request(r_i)$}
	\justifying
		\begin{algorithmic}[1]
		    \small
			\Require an incoming request in form $r_{i}=<s_{j}, t_{i}, D_{i}>$
			\Ensure the transfer cost of serving this request
			\State transfer cost $C_T \leftarrow 0$
			\State $\Delta t \leftarrow \frac{\lambda}{\mu}$
			\If{(request $r_i$ arrives $s_j$ at time $t_i$)}
				\If{(request $r_i$ contains a single data item $d_i$)}
					\If{($E[d_i]^j = 0$)}
						\State $r_i$ \mr{served by} a transfer \mr{from} $s_k$ with $d_i$, $k \neq j$; 
						\State $C_T \leftarrow C_T + \lambda$
						\State $E[d_i]^j \leftarrow t_i + \Delta t$
						\State $c[d_i] \leftarrow c[d_i] + 1$
					\ElsIf{($E[d_i]^j \neq 0$)}
						\State serve $r_i$ by the local copy on $s_j$
						\State $E[d_i]^j \leftarrow t_i + \Delta t$
					\EndIf
				\State $r_{<i>}^j \leftarrow t_i$
				\ElsIf{(request $r_i$ contains two data items $d_{i1}$ and $d_{i2}$)}
					\If{($E[d_{i1}^j] = 0$ and $E[d_{i2}^j] = 0$)}
						\If{($d_i = (d_{i1}, d_{i2}) \in FreqI$)}
							\State $d_{i1}$ \& $d_{i2}$ are transferred \mr{in two ways for} cost minimisation: \mr{1)} individual item from any server that caches a copy;
							\mr{2)} packed items from any server with both.
							\If{transferred individually}
								\State $C_T \leftarrow C_T + 2 \lambda$
							\Else
								\State $C_T \leftarrow C_T + 2 \alpha \lambda$
							\EndIf
						\Else
							\State $d_{i1}$ \& $d_{i2}$ \mr{trans. }them from $s^k$, $k \neq j$; 
							\State $C_T \leftarrow C_T + 2 \lambda$
						\EndIf
						\State $E[d_{i1}^j], E[d_{i2}^j] \leftarrow t_i + \Delta t$
						\State $c[d_{i1}] \leftarrow c[d_{i1}] + 1$, $c[d_{i2}] \leftarrow c[d_{i2}] + 1$
					\ElsIf{($E[d_{i1}^j] \neq 0$ and $E[d_{i2}^j] \neq 0$)}
						\State serve request $r_i$ by local copy on $s_j$
						\State $E[d_{i1}^j], E[d_{i2}^j] \leftarrow t_i + \Delta t$
					\Else
						\State $d_{i1}$ or $d_{i2}$ is \mr{missed} locally, \mr{trans. }from $s_k$, $k \neq j$; 
						\State $C_T \leftarrow C_T + \lambda$
						\State $E[d_{i1}^j], E[d_{i2}^j] \leftarrow t_i + \Delta t$
						\State $c[d_{i*}] \leftarrow c[d_{i*}] + 1$, $d_{i*}$ is the absent data item
					\EndIf
					\State $r_{<i1>}^j, r_{<i2>}^j \leftarrow t_i$
				\EndIf
			\EndIf
			\State \Return $C_T$
		\end{algorithmic}
\end{breakablealgorithm}

The request serving mechanism of the \emph{PackCache} algorithm has been shown in Algorithm \ref{alg:serve_request}. Upon receiving a single data item request (\D{line 4 - 14}), the request serving component will transfer the requested data item from any server $s_k$ who possesses an alive copy if the requested data item is not locally cached on server $s_j$, i.e., $E[d_{i}]^j = 0$. Tracker $E[d_i]^j$, $c[d_i]$ and the transfer cost $C_T$ will be updated accordingly as shown in \D{line 7 - 9}. On the other hand, if the requested data item is available locally, it will be served directly without transferring as shown in \D{line 10 - 12}. Finally, tracker $r_{<i>}^j$ \mr{is} updated to reflect the recent request \mr{for} the data item. 

In the \emph{PackCache} setting, requests with two data items are also allowed. The request serving works similarly with the single data item request case, except in the transfer scenarios. As indicated in \D{line 17 - 23}, when both requested data items are absent, a discounted transfer cost can be applied if these two data items form an itemset that is frequent, otherwise they will be served in an individual manner and the usual transfer cost will be applied as in \D{line 24 - 26}. Depending on the length of the idle caching time, the algorithm will use the most cost-efficient way to perform the transfer, either transfer these two data items individually, or transfer them in a packed manner. On the other hand, if both requested items are cached locally, no extra transfer is required and the request will be served directly as in \D{line 30 - 32}. Similar to the single data item requests, as shown in \D{line 33 - 37}, when any one of the requested data item is not stored locally, the transfer of the lacked data item is done individually, making the discount not applicable. Note that in this case, as well as the single data item request case, the \emph{PackCache} algorithm will only transfer the requested data item that is locally absent, instead of transferring a data item package which contains the required data item. Since in the online setting, complete knowledge about the entire request sequence is lacked, therefore it may not be worthy to use a data item package to satisfy a single data item request, which distinguishes the \emph{PackCache} algorithm with its offline counterpart. Finally, tracker $r_{<i>}$ \mr{is} updated. 

\begin{breakablealgorithm}\label{alg:copy_expire}
	\caption{Function $copy\_expire(d_p, s_j, t_i))$}
	\justifying
		\begin{algorithmic}[1]
		    \small
			\Require the copy expiration event information
			\Ensure no output will be yielded by this function
			\State $\Delta t \leftarrow \frac{\lambda}{\mu}$
			\If{($c[d_p] = 1$)}
				\State $E[d_p]^j \leftarrow t_i + \Delta t$
			\Else
				\State drop the $d_p$ copy at $s_j$
				\State $E[d_p]^j \leftarrow 0$
				\State $c[d_p] \leftarrow c[d_p] - 1$
			\EndIf
		\end{algorithmic}  
\end{breakablealgorithm}

For the copy expiration handler, we adopt the mechanism inspired by the \textit{anticipatory caching} concept. The period $\Delta t = \lambda / \mu$ is calculated as shown in \D{line 1} in Algorithm \ref{alg:copy_expire} and is also used in Algorithm \ref{alg:serve_request}. If the period between the current time and the time of the latest previous request of this data item is less than or equal to $\Delta t$, this data item is worth being cached locally as the caching cost is less than or equal to the transfer cost incurred by the eviction of this data item after its latest previous request. When the period $\Delta t$ is reached, if this data item has more than one alive copies, then the local copy will be dropped for caching cost efficiency as in \D{line 4 - 7} in Algorithm \ref{alg:copy_expire}. Otherwise, if the local copy is the only alive copy, then its expiration time will be extended by another $\Delta t$ to prevent data loss as in \D{line 2 - 3} in Algorithm \ref{alg:copy_expire}.

\subsection{Competitive Analysis}
\label{sec:online_packcache_algorithm_competitive_analysis}

The design of the \emph{PackCache} algorithm produces the following observations: 

\textbf{Observation 1.} \textit{If $c[d_k] > 1$, i.e., there exists more than one alive copies of data item $d_k$, then no copy can survive for more than $\Delta t$ on any server $s_j, j \in [1, m]$. }

\textbf{Observation 2.} \textit{Since the algorithm has no knowledge about future requests, in the worst case, it needs to cache $d_k$ after satisfying its request for $\Delta t$ period. However, if no subsequent request of $d_k$ comes in $\Delta t$ period, the optimal algorithm will not cache it at all. }

\textbf{Observation 3.} \textit{Data loss will not happen, hence at any given time, there is always a copy that can serve the incoming request, either by local caching or by transferring. }

Based on these observations, we give the following theorems of the competitive ratio of the \emph{PackCache} algorithm: 

\begin{theorem}
	The \emph{PackCache} algorithm is $\frac{2}{\alpha}$-competitive. 
\end{theorem}

\begin{proof}
	In the proof, we denote $C_{On\_PC}^i$ as the cost of serving data request $r_i$ by the proposed \emph{PackCache} algorithm, and denote $C_{OPT}^i$ as the cost of serving the same data request by the optimal offline way in the packed setting. We will first discuss the proof of single data item requests, then extend the proof to double data item requests. 

	Considering the following cases for a single data item request:
	
	\textbf{Case 1}: If $r_i = <s_j, t_i, \{d_k\}>$ is the first request that arrives at server $s_j$, $j \in [1, m]$, then we have $C_{On\_PC}^i = \dt{\mu \Delta t + \lambda = 2 \lambda}$ and $C_{OPT}^i = \lambda$. Hence we have $\frac{C_{On\_PC}^i}{C_{OPT}^i} = 2 \leq \frac{2}{\alpha}$, given that $0 < \alpha \leq 1$. 

	\textbf{Case 2}: When the request $r_i = <s_j, t_i, \{d_k\}>$ arrives, two scenarios should be considered as follows: 

	\textbf{Case 2.1}: If $t_i \in [r_{<k>}^j, r_{<k>}^j + \Delta t]$ and $E[d_k]^j \neq 0$, which means there is a local copy ready to serve the request and the caching period is within $\Delta t$, therefore we have $C_{On\_PC}^i = \mu \Delta t \le \lambda$, $C_{OPT}^i = \mu \Delta t \le \lambda$, hence $\frac{C_{On\_PC}^i}{C_{OPT}^i} = 1 < \frac{2}{\alpha}$. 

	\textbf{Case 2.2}: If $t_i > r_{<k>}^j + \Delta t$ and $E[d_k]^j = 0$, i.e., there exists no local cache of the data item to server the request. Then the transfer will occur, i.e., $C_{On\_PC}^i = \mu \Delta t + \lambda = 2 \lambda$, $C_{OPT}^i = \lambda$, hence we have $\frac{C_{On\_PC}^i}{C_{OPT}^i} = 2 \leq \frac{2}{\alpha}$. 

	Now lets extend the above proof of single data item request to double data item request. For the double data item requests, if these two data items belong to a frequent itemset, a discount factor $0 < \alpha \leq 1$ can be applied during data item transferring, otherwise, the discount factor is not applicable. Firstly, for those double data item requests in which the two data items cannot form a frequent itemset, they will be served in a separated manner and the discount factor is not applicable. Therefore, the cost in this case is twice in the analysis of a single data item request. The constant $2$ will be cancelled out for both the \emph{PackCache} and the optimal algorithm and hence the competitive ratio result remains unchanged. On the other hand, for those double data item requests in which the two data items can form a frequent itemset, the result of Case $2.1$ will remain unchanged. In that case, the request is served using local cached copy and hence the discount factor is not involved and the result is not affected. 

	For Case $1$, it needs to be reconsidered under the double data item request: 

	\textbf{Case 1'}: We have $C_{On\_PC}^i = 2 \mu \Delta t + 2 \lambda = 4 \lambda$. On the other hand, $C_{OPT}^i = 2 \alpha \lambda$. Hence, $\frac{C_{On\_PC}^i}{C_{OPT}^i} = \frac{2}{\alpha}$. 

	And for Case $2.2$, it also needs to be reconsidered in two scenarios as follows: 

	\textbf{Case 2.2'-1}: If neither of these two requested data item has available local cache copy, i.e., $E[d_{k1}]^j = E[d_{k2}]^j = 0$. Then, we have $C_{On\_PC}^i = 2 \mu \Delta t + 2 \lambda = 4 \lambda$, while $C_{OPT}^i = 2 \alpha \lambda$. Hence, we have $\frac{C_{On\_PC}^i}{C_{OPT}^i} = \frac{2}{\alpha}$. 

	\textbf{Case 2.2'-2}: If only one of the requested data item is absent locally, then we only need to transfer that data item individually. Hence, we have $C_{On\_PC}^i = \mu \Delta t + \lambda = 2 \lambda$, while $C_{OPT}^i = \lambda$. Therefore, we have $\frac{C_{On\_PC}^i}{C_{OPT}^i} = 2 \leq \frac{2}{\alpha}$, given that $0 < \alpha \leq 1$. 

	Since the $C_{On\_PC}^i$ is the cost of serving request $r_i$, same for $C_{OPT}^i$ which is the cost of the corresponding optimal algorithm, therefore, for the entire request sequence $R, |R| = n$,  
	\begin{equation}
		\frac{C_{On\_PC}^R}{C_{OPT}^R} = \frac{\sum_{1 \le i \le n} C_{On\_PC}^i}{\sum_{1 \le i \le n} C_{OPT}^i}
	\end{equation}

	Hence, we can conclude the theorem with the following result: 
	\begin{equation}
		\lim_{n \rightarrow + \infty} \frac{C_{On\_PC}^R}{C_{OPT}^R} = \frac{2}{\alpha}
	\end{equation}

\end{proof}

The following theorem shows that there does not exist any deterministic online algorithm that can yield a better performance than $\frac{2}{\alpha}\times$ optimal \mr{result}, implying our algorithm is tight.

\begin{theorem}
	The competitive ratio of the online packable caching problem is lower bounded by $\frac{2}{\alpha}$. 
\end{theorem}

\vspace{-0.3cm}

\begin{proof}
    Without loss of generality, we establish a special instance with two data items in which both $\lambda$ and $\mu$ are set to $1$, \mr{and thus} $\Delta t$ is $1$. Initially, at least one server has both data items, and at least one server has each single data item. We define $l_i$ to be length of local caching after satisfying request $r_i$. As mentioned in the previous proof, $2/\alpha$ will only appear in Case 1 and 2.2'-1, where both requested data items are absent locally. Hence, we only focus on the data pair request in which both items are absent. 
    
    We prove the theorem by reduction. Initially, a request $r_1 = <s_i, 0, \{d_{1}, d_{2}\}>$ arrives, both data items are absent and hence the request will be satisfied by a transfer. Then, the local caching length $l_1$ will have the following two cases: 
    
    \textbf{Case 1.1}: If $l_1 = 1$, then there is no subsequent requests comes after $r_1$, which yields $\frac{\mathcal{A}}{OPT} = \frac{2 + 2 l_1}{2\alpha} = \frac{2}{\alpha}$. $l_1 > 1$ will never hold since the copy will expire. 
    
    \textbf{Case 1.2}: If $l_1 < 1$, it means there is a request comes in $\epsilon$ time, $\epsilon < 1$. Similarly, we have $l_2$ to be the caching length after satisfying $r_2$, and hence it leads to two cases similar with the above:
    
    \textbf{Case 2.1}: If $l_2 = 1$, then there is no subsequent requests comes after $r_2$, which yields $\frac{\mathcal{A}}{OPT} = \frac{2 + 2 (l_1 + l_2)}{2\alpha + 2 min(\alpha, l_1)} < \frac{2}{\alpha}$.
    
    \textbf{Case 2.2}: If $l_2 < 1$, it repeats Case 1.2. We keep receiving request $r_k$. Assume upon request $r_{k-1}$, $\frac{\mathcal{A}^{k-1}}{OPT^{k-1}} \leq \frac{2}{\alpha}$ holds, we have the following:
    
    \textbf{Case k}: We have $\frac{\mathcal{A}^k}{OPT^k} = \frac{2 + 2 l_k}{\alpha + 2 min(\alpha, l_k)} < \frac{2}{\alpha}$ and it holds when $l_k \leq 1$ so we omit Case k.2. Hence, we \mr{conclude} that $2/\alpha$ is a lower bound of the competitive ratio, \mr{implying that} no deterministic algorithm can do better than this. 
\end{proof}

\begin{spacing}{0.8}
\subsection{Complexity Analysis}
\label{sec:complexity_analysis}
\end{spacing}

The implementation of the algorithm follows the workflow described in Algorithm \ref{alg:online_packcache_algorithm} - \ref{alg:copy_expire} in Section \ref{sec:an_online_2_approximation_algorithm}. In terms of the space complexity, maintaining $E[d_k]^j$ costs the highest space consumption among all tracker variables and is $O(mn)$, while constructing the FP-Tree has a space complexity $O(n^2)$. Given that the number of servers $m$ is far less than the number of requests $n$, therefore, the overall space complexity of the \emph{PackCache} algorithm is $O(n^2)$. 

As for the time complexity, when serving each incoming data request, manipulating $E[d_k]^j$, $c[d_k]$ and $r_{<i>}^j$ can be done using $O(1)$ time with efficient implementation. Hence, serving request in general remains in constant time complexity, which is highly efficient. Despite that constructing the FP-Tree, handling expired copies and calculating the caching cost may cause some overhead, fortunately, all these operations can be processed by background daemons running in parallel with the request serving process. Hence, they will not impair the time complexity of request serving.


\section{Performance Studies}\label{sec:simulation_studies}

To verify the performance of our algorithm in practice, extensive experiments are conducted. We design a solver in Python, which effectively implements our algorithm. Follow \cite{huang2019dp_greedy}, the experiment data comes from the taxi trace data from City of Shenzhen in China. The territory of city is partitioned into 50 parts, each maintains a cache server to serve the user requests of taxis, which are regarded as shared data items. For instance, at time $t_i$, server in region $s_j$ receives a request of two taxis will be used as a request in our experiment. The dataset contains various request pairs with relatively high Jaccard Similarity. According to the research results~\cite{huang2019dp_greedy}, the trace of the taxi can be roughly seen as the trace on how data are requested from different servers. 

The algorithm is characterised by several parameters, which include number of data items $k$, number of caching nodes $m$, number of requests $n$, the discount factor $\alpha$, the minimum support threshold $\gamma$ for the FP-Tree frequent data itemset mining, the caching cost $\mu$ and the transfer cost $\lambda$. To concentrate our study on the factors we concerned about, we deliberately ignore some other factors that may influence the algorithm, such as CPU power, network condition and bandwidth of the network, etc. On the other hand, we take the average cost as the major performance metric since many other performances can be reflected from it such as the network bandwidth occupancy rate. 

During experiments, $10$ taxis are randomly selected, each acts as a distinct data item ($d_{1}, d_{2}, \cdots, d_{10}$) as this value can be well handled and without loss of generality to reflect general case. We partition the city into $50$ parts, each having a caching server, and set the discount factor $\alpha = 0.8$. The default transfer cost $\lambda$ and caching cost $\mu$ are all set to be $3$ to balance between transferring and caching. We set the minimum support $\gamma = 0.01$ based on our experience on research of human mobility behaviors in metropolitan city~\cite{huang2019dp_greedy}. Finally, several factors are varied to testify the effectiveness and robustness of the \emph{PackCache} algorithm. 

When evaluating the \emph{PackCache} algorithm, we compare it with the algorithm that individually serves packed requests without the data packing mechanism to demonstrate the effectiveness of the packable \emph{PackCache} algorithm. \rw{Besides, we also compare with the offline counterpart to show that although the performance is lower bounded by $2/\alpha$, the \emph{PackCache} algorithm usually performs better than the theoretical lower bound. }

\subsection{Impact of Ratio $\rho = \lambda / \mu$}
\label{sec:impact_on_ratio_rho_on_online_packcache_algorithm}
\vspace{-0.08cm}

\begin{figure*}[!ht]
	\setlength{\abovecaptionskip}{0pt}
	\setlength{\belowcaptionskip}{-15pt}
	\centering
	\includegraphics[scale=0.3]{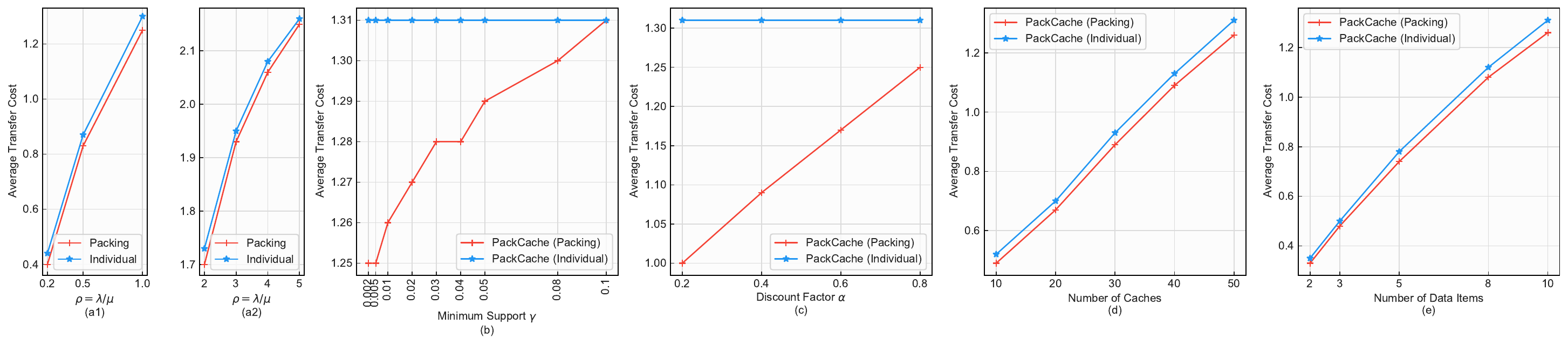}
	\caption{The average transfer cost of \emph{PackCache} algorithm and its individually-served counterpart under (a1) \& (a2) different $\rho$ ratios, (b) different minimum support $\gamma$, (c) different discount factor $\alpha$, (d) different number of cache servers and (e) different number of data items. }
	\label{fig:combined}\vspace{-6mm}
\end{figure*}

When evaluating the effectiveness and robustness of the \emph{PackCache} algorithm \mr{with} different $\rho$ \mr{ratios}, we set $\lambda + \mu = 6$ intentionally. The experimental results of the transfer cost is illustrated in Fig. \ref{fig:combined}(a). 

According to Fig. \ref{fig:combined}(a), the average transfer cost increases as the ratio $\rho$ increases. It is natural to observe because as the ratio $\rho$ increases, the transfer cost $\lambda$ is gradually emphasised while the caching cost $\mu$ is gradually declined. Under all $\rho$ settings ranging between $0.2$ and $5.0$, the \emph{PackCache} algorithm outperforms its individually-served counterpart, which demonstrates the superiority and robustness of the \emph{PackCache} algorithm. For instance, when $\rho$ is set to $1$, the cost reduction achieved by the \emph{PackCache} is around $4.6\%$. Besides, we can observe that the cost does not increase in a linear trend as the $\rho$ ratio raises. The reason is that when the ratio $\rho$ raises, the transfer cost becomes higher than the caching cost, which gradually encourages local caching of data items instead of transferring them to serve the requests. Data items will be cached locally for longer periods, the local caching becomes less frequent to expire, leading to less transfers being performed. 

\begin{spacing}{0.8}
\subsection{Impact of Minimum Support $\gamma$}
\label{sec:impact_of_minimum_support_phi_on_online_packcache_algorithm}
\end{spacing}

The effectiveness of the \emph{PackCache} algorithm under different minimum support $\gamma$ during FP-Tree construction has been illustrated in Fig. \ref{fig:combined}(b). The minimum support value controls the threshold of whether two data items will be considered frequent or not. The higher the $\gamma$ is, the more frequently should both data items be co-utilised in order to be considered as frequent. A higher minimum support $\gamma$ will discourage the discounted packing from being applied and vice versa. Hence, by observing Fig. \ref{fig:combined}(b), we notice that the average transfer cost of \emph{PackCache} algorithm increases as the minimum support raises. The average transfer cost even approaching its individually-served counterpart when the minimum support becomes relatively large, i.e., $0.1$ in this case. This is natural to observe since when the minimum support $\gamma$ is relatively small, more co-utilised data items will be considered as being frequent and hence the benefit brought by the data packing can be fully exploited. While the minimum support $\gamma$ becomes higher, less data co-utilisation will be considered to be frequent and hence the data packing benefit gradually diminishes. Therefore, the performance of the \emph{PackCache} algorithm will approach its individual counterpart as the minimum support increases. 

Specifically, when the minimum support $\gamma$ is set to the default value $0.01$, the cost reduction achieved by the \emph{PackCache} algorithm is around $3.8\%$. Furthermore, under all minimum support settings except the extremely high setting, a significant cost reduction has been observed, which demonstrates the effectiveness of the \emph{PackCache} algorithm and its robustness in terms of varied $\gamma$ settings. 

\vspace{0.05cm}
\begin{spacing}{0.9}
\subsection{Impact of Discount Factor $\alpha$}
\label{sec:impact_of_discount_factor_alpha_on_online_packcache_algorithm}
\end{spacing}

The performance of \emph{PackCache} algorithm under different discount factor $\alpha$ is shown in Fig. \ref{fig:combined}(c). The discount factor $\alpha$ controls the benefit that the packing mechanism can bring. The higher the discount factor, the less benefit when utilising the packing mechanism. Hence, as observed from Fig. \ref{fig:combined}(c), the average transfer cost increases linearly as the discount factor $\alpha$ raises. Despite the raise of the average transfer cost, the \emph{PackCache} algorithm still outperforms its individual counterpart by a large margin under all $\alpha$ settings. For instance, when $\alpha$ is $0.6$, the cost reduction achieved by the \emph{PackCache} algorithm is $10.7\%$. Even for the highest setting $0.8$, which is the default value we utilised following \ref{sec:simulation_studies}, the cost reduction is still around $4.6\%$. Hence, the results demonstrates the excellent performance and robustness of the \emph{PackCache} algorithm. 

\vspace{0.05cm}
\subsection{Scalability of the Algorithm}
\label{sec:scalability_of_online_packcache_algorithm}

\begin{figure}[!ht]
	\setlength{\abovecaptionskip}{0pt}
	\setlength{\belowcaptionskip}{-15pt}
	\centering
	\includegraphics[scale=0.27]{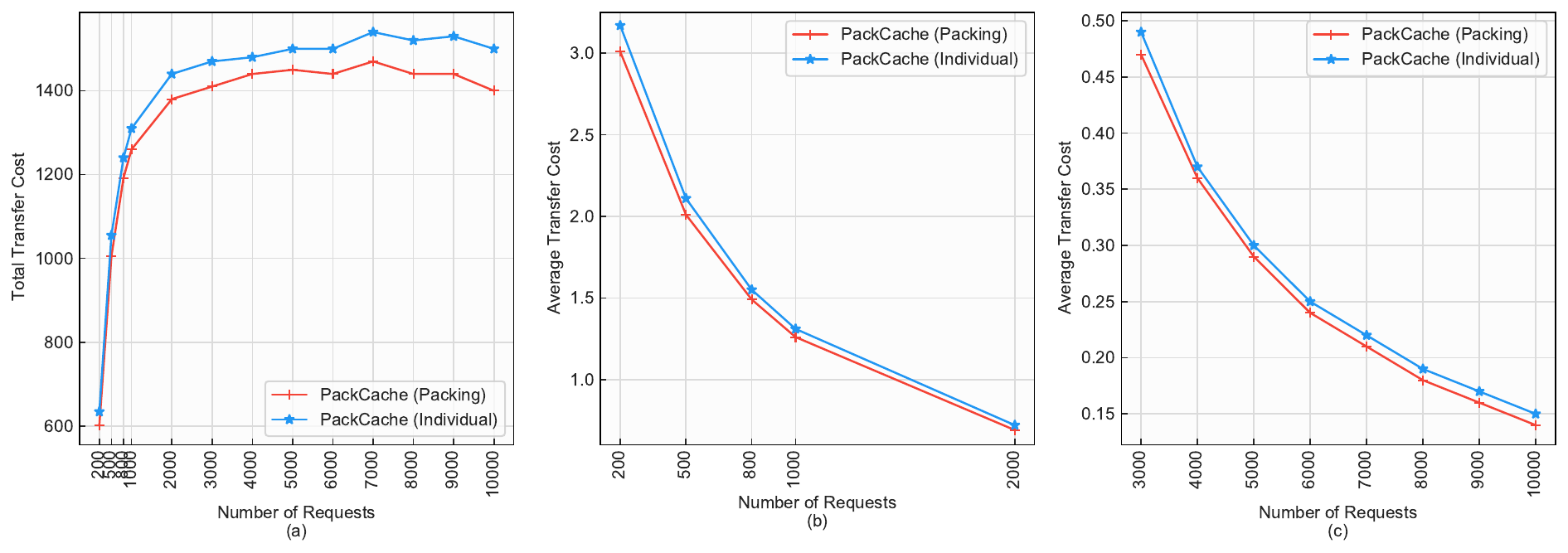}
	\caption{The total and average transfer cost of the \emph{PackCache} algorithm and its individual counterpart when serving different number of requests. }
	\label{fig:requests_transfer_online}\vspace{-2mm}
\end{figure}

To testify the scalability of the \emph{PackCache} algorithm, three variables are adjusted, i.e., number of requests, number of cache servers, and number of data items. The evaluation results are presented in Fig. \ref{fig:requests_transfer_online}, \ref{fig:combined}(d) - \ref{fig:combined}(e), respectively. 

\textit{Scalability (Number of requests)}: From Fig. \ref{fig:requests_transfer_online}(a), it is natural that the total transfer cost of request serving rises when the number of requests increases from $200$ to $10000$. The total transfer cost reaches a plateau when the number of requests becomes even larger. The reason is that when the number of requests becomes larger, the requests will come in a denser manner, making the cached copies hardly expire. Hence, it significantly reduces the need of copy transfer, and hence it also results in the transfer cost stops growing. 

In terms of average request transfer cost, a clear decreasing trend is observed from Fig. \ref{fig:requests_transfer_online}(b) and (c). For instance, when the number of requests is $500$ and $2000$, the \emph{PackCache} algorithm achieves cost reduction of $4.7\%$ and $4.2\%$, respectively, which verifies the scalability of the \emph{PackCache} algorithm when serving varied number of requests. 

\textit{Scalability (Number of cache servers)}: The performance of the \emph{PackCache} algorithm when working under different number of cache servers has been presented in Fig. \ref{fig:combined}(d). The average transfer cost increases linearly when the number of cache servers raises. The linear trend indicates that the \emph{PackCache} algorithm scales stably when the number of cache servers varies, demonstrating excellent scalability in terms of different number of cache servers. Under different number of cache servers, the \emph{PackCache} algorithm achieves a cost drop around $4.3\%$, which further verifies the effectiveness of the \emph{PackCache} algorithm. 

\textit{Scalability (Number of data items)}: Finally, the \emph{PackCache} algorithm is evaluated under different number of shared data items. From Fig. \ref{fig:combined}(e), we can observe a relatively stable increasing trend of the average transfer cost when the number of data items increases. Under all settings, the \emph{PackCache} algorithm constantly achieves superior performance. More specifically, cost saving of $5.7\%$, $5.1\%$ and $3.8\%$ are achieved when the number of data item is set to be $2$, $5$ and $10$, respectively. The reason why the cost reduction gradually decreases is that as the number of data items increases, more double data item pairs can be formed, hence, less pairs become frequent and hence the packing mechanism is less frequently applied. Therefore, as the number of data items keeps growing, the performance of the \emph{PackCache} algorithm will gradually approach its individually-served counterpart, as the benefit brought by the packing mechanism will gradually diminish. However, when the number of data items are relatively small, we can observe significant cost reduction, which verifies the effectiveness and scalability of the \emph{PackCache} algorithm. 

\subsection{\rw{Comparison with Offline Algorithm}}
\label{sec:comparison_with_offline_algorithm}

\rw{To demonstrate the effectiveness, we also compare the \emph{PackCache} algorithm with its offline counterpart. The cost yielded by the \emph{PackCache} algorithm is $1.04$, $1.41$, and $1.31$ times higher than its offline counterpart when $\rho$ varies between $0.5$ and $2$. When $\alpha$ varies from $0.6$ to $0.8$, the cost is $1.13$ and $1.41$ times higher than the offline version. The \emph{PackCache} achieves costs that are $1.04$, $1.27$ and $1.41$ times higher than its offline counterpart in settings with $10$, $30$ and $50$ servers, and are $1.12$ and $1.41$ times higher in settings with $5$ and $10$ data items. Finally, $1.5$ and $1.41$ times higher cost are produced when the number of requests varied from $500$ to $1000$. As we can observe, although the performance of \emph{PackCache} is lower bounded by $2/\alpha$ ($1.6$ under default $\alpha = 0.8$), in various settings, this theoretical lower bound is usually not reached, which demonstrates the effectiveness of the \emph{PackCache} algorithm. }
\vspace{1mm}
\begin{spacing}{0.8}
\subsection{\rw{Performance under Real Cost Model}}
\label{sec:performance_under_real_cost_model}
\end{spacing}

We use the real service prices to demonstrate that \emph{PackCache} algorithm can achieve a significant cost saving under real deployment. Google Cloud charges $\$0.04$ and $\$0.08$ for caching and transferring a GB of data. We suppose the algorithm daily serves $1000$ requests for total $1000$GB of data items. \mr{Given a reasonable range of discount factor between $0.6$ and $0.9$ depending on the selected compression technique, we use $\alpha$ of $0.6$ and $0.8$ as representative scenarios. The system can yield a yearly cost saving of $\$1306.7$ out of $\$11468$ and $\$449$ out of $\$12002$, which is approximately equivalent to reducing $21778$ and $7483$ GB of data being handled, respectively. Hence, the \emph{PackCache} algorithm is beneficial in real settings. }


\section{Conclusion}
\label{sec:section_conclusion}

In this paper, we studied \mr{a} data caching problem in the cloud with cost minimisation as the \mr{goal}. Given that serving data requests in a packable manner is usually more cost effective than in its individual counterpart (i.e., non-packing), we are among the first to propose a \mr{time-space efficient} \emph{PackCache} algorithm, which leverages FP-Tree to mine frequently co-utilised data in an online setting and exploits the concept of anticipatory caching for service cost reduction. We showed the algorithm is $2/\alpha$ competitive with respect to a homogeneous cost mode, reaching the lower bound of the competitive ratio for any deterministic online algorithm on \mr{this problem}. Finally, we evaluated the performance of the algorithm via experimental studies to show its actual cost-effectiveness and scalability \mr{in practice}.

\vspace{2mm}
\begin{spacing}{0.85}
\bibliographystyle{IEEEtran}
\bibliography{PackCacheTC}
\end{spacing}


\begin{IEEEbiography}[{\includegraphics[width=1in,height=1.25in,clip,keepaspectratio]{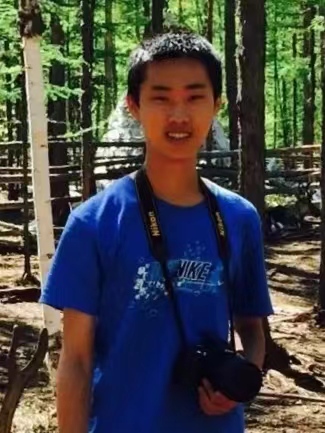}}]{Jiashu Wu} received BSc. degree in Computer Science and Financial Mathematics \& Statistics from the University of Sydney, Australia (2018), and M.IT degree in Artificial Intelligence from the University of Melbourne, Australia (2020). He is currently pursuing his Ph.D at the University of Chinese Academy of Sciences (Shenzhen Institute of Advanced Technology, Chinese Academy of Sciences). His research interests including transfer learning and cloud computing. 
\end{IEEEbiography}

\begin{IEEEbiography}[{\includegraphics[width=1in,height=1.25in,clip,keepaspectratio]{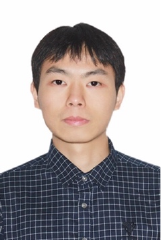}}]{Hao Dai} received the BS and M.Sc. degrees in Communication and Electronic Technology from the Wuhan University of Technology in 2015 and 2017, respectively. He is currently working toward the Ph.D. degree in the Shenzhen Institutes of Advanced Technology, Chinese Academy of Sciences. His research interests include mobile edge computing, federated learning and deep reinforcement learning. 
\end{IEEEbiography}

\begin{IEEEbiography}[{\includegraphics[width=1in,height=1.25in,clip,keepaspectratio]{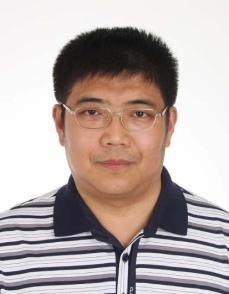}}]{Yang Wang} received the BSc degree in applied mathematics from Ocean University of China, in 1989, and the M.Sc. degree in computer science from Carleton University, in 2001, and the Ph.D degree in computer science from the University of Alberta, Canada, in 2008. He is currently in Shenzhen Institutes of Advanced Technology, Chinese Academy of Sciences, as a professor. His research interest includes cloud computing, big data analytics, and Java virtual machine on multicores. He is an Alberta Industry R\&D Associate (2009-2011), and a Canadian Fulbright Scholar (2014-2015). 
\end{IEEEbiography}

\begin{IEEEbiography}[{\includegraphics[width=1in,height=1.25in,clip,keepaspectratio]{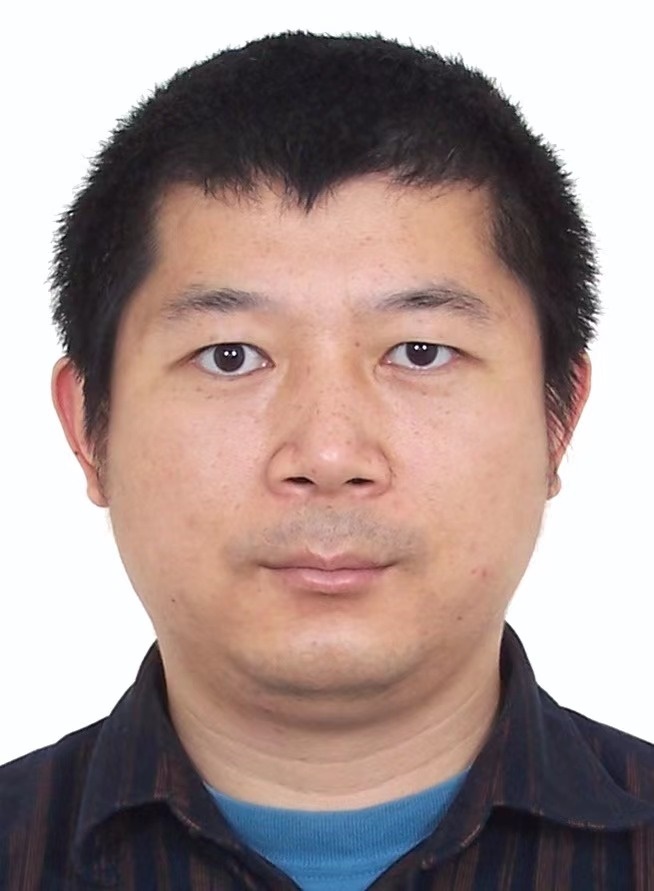}}]{Yong Zhang} received his Ph.D. in the Department of Computer Science and Engineering at Fudan University in 2007. He is now a Professor in SIAT, CAS, Honorary Professor at the University of Hong Kong. Before joining SIAT, he worked as Post-Doctoral Fellow and Senior Researcher in TU-Berlin and HKU. He has published more than 100 papers in refereed journals and conferences. His research interests include design and analysis of algorithms, combinatorial optimization, and wireless networks. 
\end{IEEEbiography}

\begin{IEEEbiography}[{\includegraphics[width=1in,height=1.25in,clip,keepaspectratio]{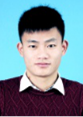}}]{Dong Huang} received the M.Eng degree from University of Chinese Academy of Sciences (Shenzhen Institute of Advanced Technology, Chinese Academy of Sciences) in 2019. His research interest includes cloud computing. 
\end{IEEEbiography}

\begin{IEEEbiography}[{\includegraphics[width=1in,height=1.25in,clip,keepaspectratio]{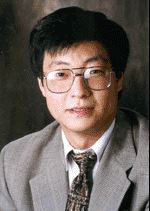}}]{Chengzhong Xu} received the Ph.D. degree from the University of Hong Kong in 1993. He is currently the Dean of Faculty of Science and Technology, University of Macau, China, and the Director of the Institute of Advanced Computing and Data Engineering, Shenzhen Institute of Advanced Technology of Chinese Academy of Sciences.His research interest includes parallel and distributed systems and cloud computing. He has published more than 400 papers in journals and conferences. He serves on a number of journal editorial boards, including IEEE TC, IEEE TPDS, IEEE TCC, JPDC and China Science Information Sciences. He is a fellow of the IEEE. 
\end{IEEEbiography}

\end{document}